\def\year{2019}\relax
\documentclass[letterpaper]{article} 
\usepackage{aaai19}  
\usepackage{times}  
\usepackage{helvet} 
\usepackage{courier}  
\usepackage[hyphens]{url}  
\usepackage{graphicx} 
\usepackage{graphicx}  
\usepackage{caption}
\usepackage{subcaption}
\usepackage{amsmath}

\title{Human self-determination within algorithmic sociotechnical systems}

\author{\Large \textbf{Bogdana Rakova, Rumman Chowdhury}\\ 
Responsible AI, Accenture\\ 
415 Mission St Floor 35\\
San Francisco, California 94105\\
\texttt{\{bogdana.rakova,rumman.chowdhury\}@accenture.com}
}

\begin{document}

\maketitle

\begin{abstract}
In order to investigate the protection of human self-determination within algorithmic sociotechnical systems, we study the relationships between the concepts of mutability, bias, feedback loops, and power dynamics. We focus on the interactions between people and algorithmic systems in the case of Recommender Systems (RS) and provide novel theoretical analysis informed by human-in-the-loop system design and Supervisory Control, in order to question the dynamics in our interactions with RSs. We explore what meaningful reliability monitoring means in the context of RSs and elaborate on the need for metrics that encompass human-algorithmic interaction. We derive a metric we call a \textit{barrier-to-exit} which is a proxy to the amount of effort a user needs to expend in order for the system to recognize their change in preference. Our goal is to highlight the assumptions and limitations of RSs and introduce a human-centered method of combating deterministic design. 
\end{abstract}


\section{Introduction}

The \textit{heart} of our research is the concept of human self-determination, self-actualization, identity, agency, and our ability to change while participating in a world where AI is ubiquitous \cite{bruni2007reassembling}. The issue of self-determination is not an external thing which can be handled or hold in hand easily. This however doesn't allow us to ignore it as some form of intellectual or technical debt \cite{JZ,sculley2015hidden}.


We analyze the co-evolution of the interaction between users and a RS in its time domain. Usually, digital footprints, logs of user interaction data over time is fed into those systems and used as "implicit" ratings. As described by \citeauthor{doi:10.1177/1359183518820366}, users stopping a video partway through, skipping over recommended items, or listening to songs multiple times are being interpreted as ratings data. A user seen through activity logs is "a ghostly presence who left traces over time" \cite{doi:10.1177/1359183518820366}. The collection of traces could never accurately represent a real person's human preferences. It is those imperfections and the consecutive errors, that lead to reduction of human self-determination.



The main contribution of our work is (1) to bring critical considerations from the field of Human Factors and Supervisory Control to question who is the entity who takes the \textit{supervisory} role in the repeated interactions between a single user and an algorithmic system, (2) to identify a gap in the evaluation metrics that are widely used to measure performance, and (3) to derive a human-algorithmic interaction metric called a \textit{barrier-to-exit} which aims to be a proxy for quantifying the ability of the AI model to recognize and allow for a change in user preference. 

We choose to focus on RSs as a target AI System in this work in order to make it more concrete and actionable, however we think that the methodologies and the metric we develop could be applied towards other kinds of AI Systems.  We hope our work will provoke meaningful discussions and inspire change in the design of AI Systems, that will ultimately positively contribute to a world where algorithmic outcomes are responsive to but cognizant of their influence on human self-determination.

\section{Related work}
Drawing from the fields of Science and Technology Studies (STS) and Anthropology, it is imperative for Computer Scientists to consider their work within a broad sociotechnical context. Engineers, Data Scientists, and AI System designers should understand the notions of algorithms as culture, institutions, ideologies, infrastructure, and actors \cite{doi:10.1177/2053951717738104,doi:10.1177/0163443716643157}. STS scholars \citeauthor{stark_hoffmann_2019} demonstrate that "the metaphors we deploy to make sense of new tools and technologies serve the dual purpose of highlighting the novel by reference to familiar, while also obscuring or abstracting away from some features" \cite{stark_hoffmann_2019}. The metaphor which we chose to depict in our analysis is that of RS as 'traps'. We build on prior work by \citeauthor{selbst2019fairness} which has outlined "five 'traps' that fair-ML work can fall into even as it attempts to be more context-aware in comparison to traditional data science" \cite{selbst2019fairness}. Ultimately, "traps offer a powerful vocabulary for articulating sociotechnical concerns "\cite{doi:10.1177/1359183518820366}. We see a close connection between the anthropological notion of traps, algorithms as infrastructure, and the emergence of self-reinforcing feedback loops in RSs, leading to the formation of filter bubbles and echo chambers \cite{Jiang:2019:DFL:3306618.3314288}. As demonstrated by \citeauthor{corsin2017spider}, "an infrastructure is a trap in slow motion. Slowed down and spread out, we can see how traps are not just devices of momentary violence, but agent of "environmentalization" \cite{corsin2017spider}, making worlds for the entities they trap". In that context, "to be caught … is to be enclosed, known, and subject of manipulation" \cite{doi:10.1177/1359183518820366}.

While there is a wealth of literature demonstrating that AI algorithms don't operate in vacuum but are instead part of complex sociotechnical systems, the debate about their consecutive unintended consequences is much more difficult and underdeveloped. \citeauthor{aragon2016developing} have identified the need for the development of a research agenda for Human-Centered Data Science employing qualitative methods to study the interdependencies between algorithms and the social context within which they operate \cite{aragon2016developing}. The recently developed fields of Critical Data Studies and Critical Data Science have studied the dynamic relationship between data and its users, through communication, sensemaking, following how datums change through time, and storytelling \cite{neff2017critique,tanweer2016impediment}. We are inspired by these recent developments in understanding data assemblages \cite{tanweer2016impediment} and choose to further investigates the longitudinal aspects of data and its users through the concept of mutability. We focus on Recommender Systems as the algorithmic actor in our analysis and also show how a similar approach could be applied in other AI-enabled systems. Recommender System algorithms are becoming ever more deeply embedded in the life of people - choosing what to experiences (movies, music, products to buy, etc), having access to employment opportunities, selecting a doctor or a hospital, choosing a drug or a treatment, or whether to participate in a medical trial, and many others. A substantial body of research has been developed in the related field of bandit problems where practitioners study sequential decision making with limited information and address the trade-off between exploration and exploitation. Some of the early work in that field by \citeauthor{ThompsonONTL}, has been motivated by clinical trials - being able to decide which treatment to use for a certain disease \cite{ThompsonONTL}.


\citeauthor{Nguyen:2014:EFB:2566486.2568012} have studied the effect of RS on Content Diversity, and demonstrated that recommendation-following users on the MovieLens platform \cite{Harper:2015:MDH:2866565.2827872} were exposed to slightly more diverse content than non-recommendation-following users, therefore "taking recommendations lessened the risk of a filter bubble" \cite{Nguyen:2014:EFB:2566486.2568012}. However the comparative analysis in their experiment excludes users who were in the middle group of sometimes following and sometimes not following recommendations. Learning from the body of work on user preferences and user behavior, we believe its critically important to include that group in the analysis.

\citeauthor{Jiang:2019:DFL:3306618.3314288} have introduced an interaction model between users and RSs to study the formation of echo chambers and filter bubbles. They develop a metric to measure the speed of degeneracy of feedback loops and study how does the system design influence it. We build on that work, by applying the interaction model they proposed in the context of the MovieLens dataset. We leverage the field of Supervisory Control to link that interaction model to the concept of mutability of user preferences. Human preferences and identity are interdependent concepts. We see user revealed preferences in the form of different kinds of user-feedback to be an imperfect proxy to a user's actual preference as well as their identity. We leverage prior work studing the ethics and consequences of online identity enforcement for users with multifaceted, changing, or non-normative identities \cite{haimson2016constructing,Haimson:2016:DFC:2858036.2858136}.

\citeauthor{herlocker2004evaluating} have developed a taxonomy of RSs evaluation metrics, including measuring regret, serendipity, novelty, the Magic Barrier and others \cite{herlocker2004evaluating}. We build on that work and demonstrate the need for a \textit{barrier-to-exit} evaluation metric. Building on the work of \citeauthor{mitchell2019model} on developing a framework for transparent model reporting, we identify (1) the need for more granular evaluation metrics for the interaction between a RS and a dynamic user identity, as well as (2) organizational changes in the RSs system design.  

\section{Mutability}
The concept of the mutability and immutability of data variables has had a wide-spread impact in the development of Computer Programming Languages. One of the foundational design paradigms in Computer Science is that of Object-Oriented Programming, originally introduced by Alan Kay in the early 70s. In object-oriented programming, an immutable object (unchangeable object) is an object whose state cannot be modified after it is created. This is in contrast to a mutable object (changeable object), which can be modified after it is created. Mutability helps Computer Scientists specify and regulate how data variables (objects) change thought the interactions between different function calls. It is one of the foundational concepts in Computer Science and object-oriented programming and without its understanding we wouldn't be able to have the AI systems of today.

From a Social Science perspective, mutability is closely connected to our ability to change, our dynamic and multifaceted human preferences, as well as multiplicity of identities. We leverage the framework of thinking of the data measurements being made in the context of Machine Learning systems as sampling from distributions which approximate unobservable mutable and immutable statistical variables. The concept of mutability is closely related to the concept of identity transitions - our ability to adapt to different life situations, such as gender transitions, divorce, career change. Identities are dynamic and mutable \cite{crown_1989,ashforth2000role,10.2307/20159762,reay2017getting} - they change and evolve over time and that allows us to develop a more coherent sense of self \cite{10.2307/j.ctt1d8h9wn}. We argue that the AI Systems we interact with should inform but not interfere with our personal sensemaking process. If they ignore the importance of mutability of data variables, RSs creators and designers, might unintentionally create systems that employ a form of algorithmic determinism which precludes our need for experimentation and exploration, and underestimate the multiplicity of our identity. 

Algorithmic Determinism is the idea that probabilistic machine learning systems can create immutable classifications of individuals, reinforcing these classifications through feedback loops, resulting in the formation of filter bubbles and so called self-fulfilling prophecies. We argue for the need for RSs design to allow for identity transitions. \citeauthor{Haimson:2016:DFC:2858036.2858136} who analyzed open-ended survey responses from 283 participants, trangender people who transitioned while active on Facebook, highlight the types of data considered problematic when separating oneself from a past identity \cite{Haimson:2016:DFC:2858036.2858136}. We look into that study as quantitative and qualitative evidence to support the case that mutability should be taken into account when designing RSs.

In what follows, we assume that:
\begin{itemize}
\item User's preferences are mutable (not constant).
\item User's implicit and explicit feedback (in terms of the log of user actions such as clicks, ratings, etc.) are an imperfect measure of the user's actual preferences. 
\item Accumulated selection, measurement and other kinds of bias will inevitably impact the way users interact with RSs.
\end{itemize} 

\section{Critical considerations from a human-in-the-loop system design perspective}

\begin{figure}[htbp]
    \centering
    \begin{subfigure}[b]{0.52\linewidth}
        \includegraphics[width=\textwidth]{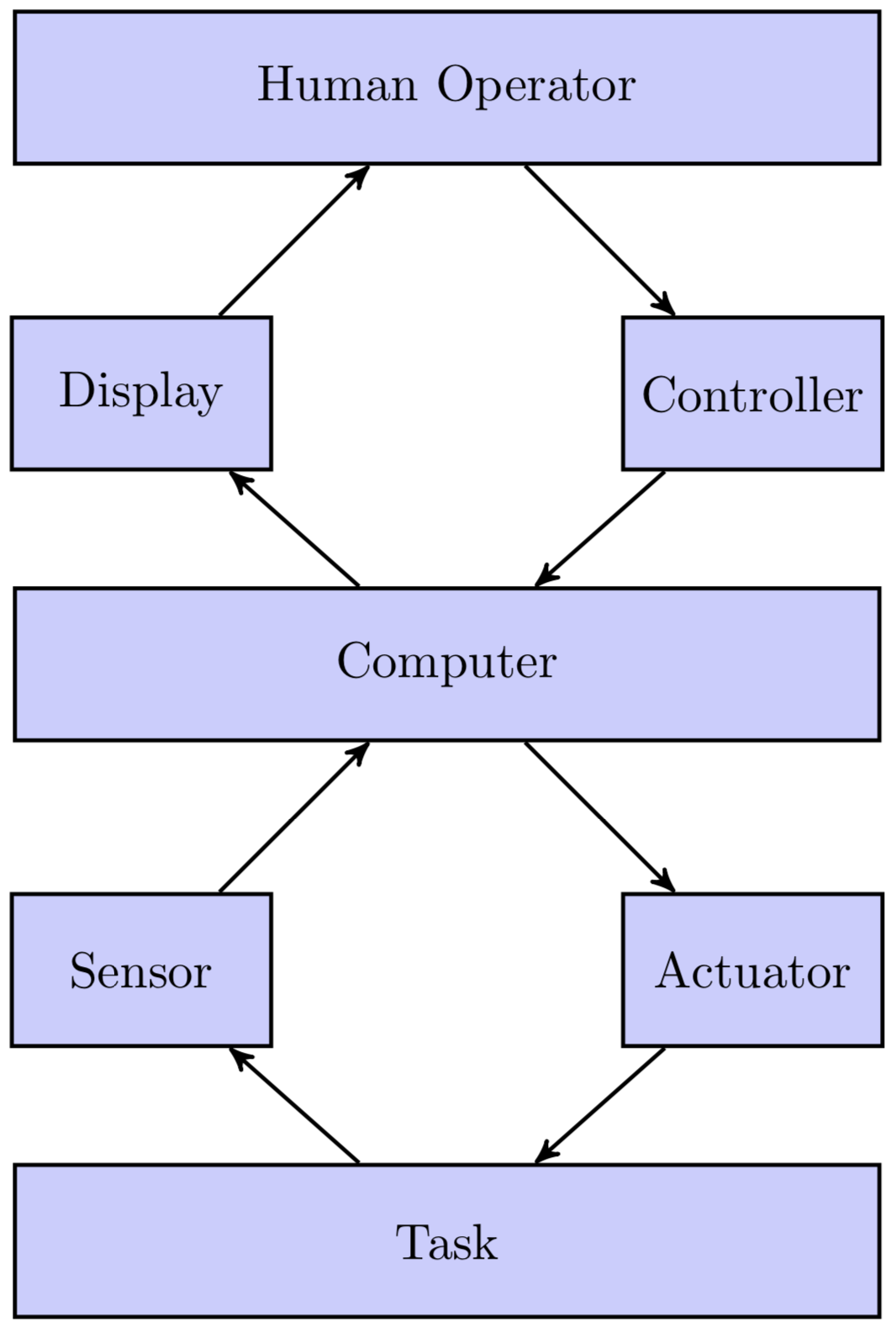}
        \caption{}
        \label{fig:setup_figure_a}
    \end{subfigure}
    \begin{subfigure}[b]{0.47\linewidth}
        \includegraphics[width=\textwidth]{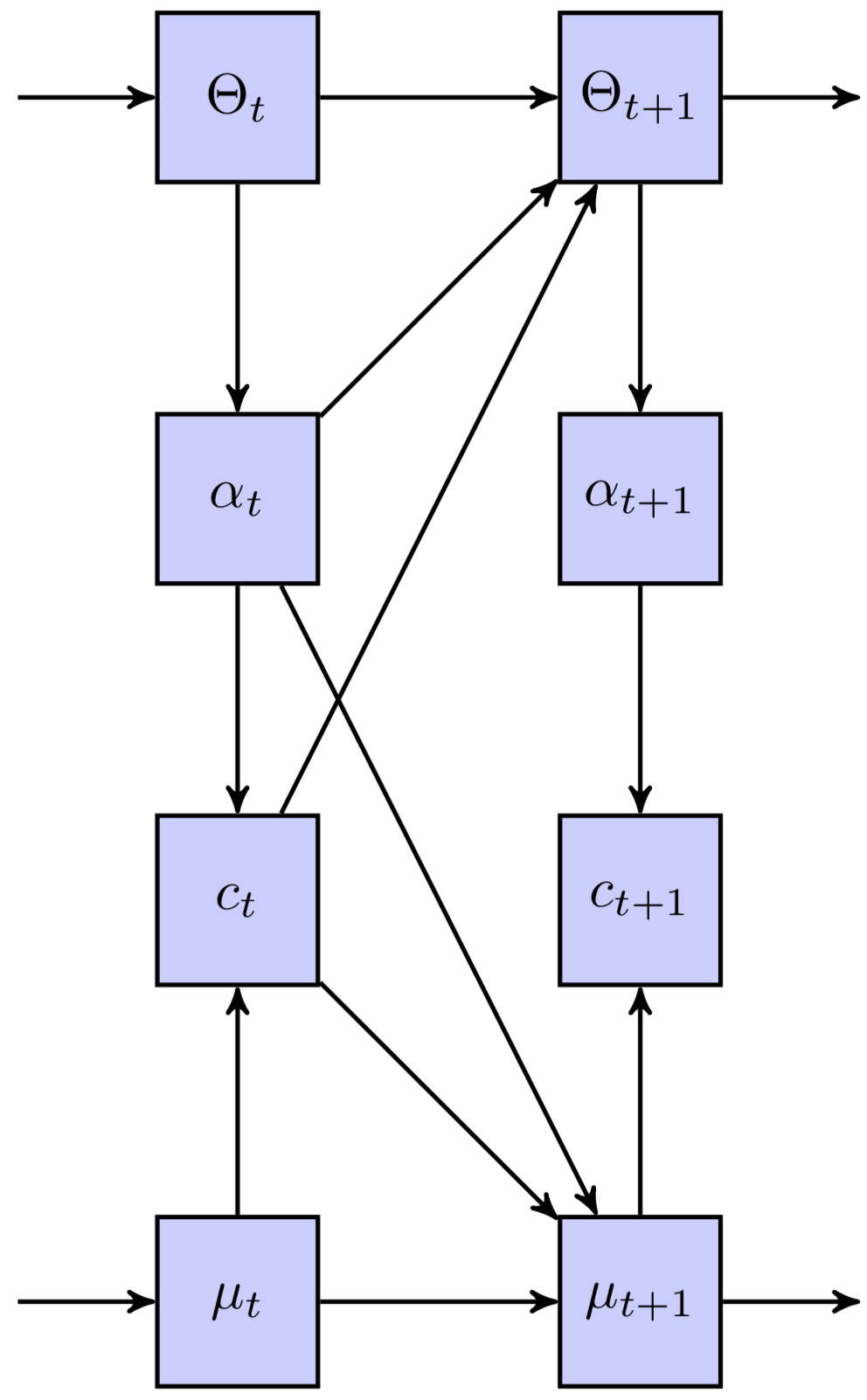}
        \caption{}
        \label{fig:setup_figure_b}
    \end{subfigure}
    \caption{(a) Human Supervisory Control flow diagram (b) Interaction model between a user and a RS over time
        \cite{Jiang:2019:DFL:3306618.3314288}.}
    \label{fig:setup_figure}
\end{figure}

In a human-in-the-loop system, a human operator is a crucial part of an automated control process, handling the tasks of supervision, exception control, optimization and maintenance. It has been studied in the theory of Supervisory Control within the field of Human Factors and Ergonomics \cite{sheridan1992telerobotics,doi:10.1002/9781118131350.ch34}. Relevant to our discussion here is specifically the concept of human supervisory control (see Figure \ref{fig:setup_figure_a}), which is a process by which "one or more human operators are intermittently programming and continually receiving information from a computer that itself closes an autonomous control loop through artificial effectors to the controlled process or task environment" \cite{sheridan1992telerobotics}. In order to understand if and how this field has underlined the kinds of AI systems we have today, it is important to understand the context out of which it emerged: "the term supervisory control is derived from the close analogy between the characteristics of a supervisor's interaction with subordinate human staff members and a person's interaction with "intelligent" automated subsystems" \cite{doi:10.1002/9781118131350.ch34}. 
In this section we first introduce some notation in order to model the interactions between a user and a RSs. We then reformulate that model in time and compare it to the human supervisory control model in order to reveal some key challenges in the design of RSs. 

We start by building on the interaction model introduced by \citeauthor{Jiang:2019:DFL:3306618.3314288} in their analysis of degenerate feedback loops in RSs (Figure \ref{fig:setup_figure_b}). We focus on the case of a single user interacting with a RS model and leave the case of multi-user interactions for future work. At every time step $t$, the model serves $l$ items to a user from a finite or countably infinite list of items $M$. Given the recommended items at time step $t$ as $\alpha_t$=($\alpha_t^1$,...,$\alpha_t^l$) $\in$ $M^l$, the user provides feedback (in terms of clicks) $c_t$ based on their current user preferences $\mu_t$=($\mu_t$($a_t^1$),...,$\mu_t$($a_t^l$)). The objective of the RS model $\Theta_t$ at time step $t$ is to provide items which the user is ultimately interested in. At the next time step, $t+1$, the model $\Theta_{t+1}$ will be updated accordingly, to take into account $a_t$ as well as $c_t$ (see Figure \ref{fig:setup_figure_b}). In their work, \citeauthor{Jiang:2019:DFL:3306618.3314288} have shown that a feedback loop exists and the user's interaction with the RS may change their interest in different items for the next interaction. The user interest $\mu_{t+1}$ may be influenced by $\mu_t$ as well as the previously recommended items $a_t$ and the previous user feedback $c_t$. They measure change in user's interest by quantifying the speed of degeneration of that feedback loop and explore degeneration at the cost of (1) compromising the model accuracy, (2) increasing the amount of exploration through randomness, as well as (3) a growing candidate pool of items recommended to the user. By monitoring of implicit and explicit user preferences through user's interactions with the RS, system designers could track the degeneracy of feedback loops and "slow them down" (remedy and mitigate the formation of filter bubbles and echo chambers) through different strategies such as those explored by \citeauthor{Jiang:2019:DFL:3306618.3314288}. 


We now leverage the field of Supervisory Control and human-in-the-loop systems in order to demonstrate that it is beneficial for AI system designers to create interaction interfaces which allow for mutability. Figure \ref{fig:setup_figure_a} represents a supervisory control diagram as defined in the field of Human Factors and Ergonomics \cite{doi:10.1002/9781118131350.ch34}. It has been studied and applied broadly to vehicle control (aircraft and spacecraft, ships, and undersea vehicles), continuous process control (oil, chemicals, power generation), and robots and discrete tasks \cite{doi:10.1002/9781118131350.ch34}. 

The human supervisor's roles are "(1) planning offline what task to do and how to do it; (2) teaching (or programming) the computer what was planned; (3) monitoring the automatic action online to make sure that all is going as planned and to detect failures; (4) intervening, which means the supervisor takes over control after the desired goal state has been reached satisfactorily, or interrupts the automatic control in emergencies to specify a new goal state and reprogram a new procedure; and (5) learning from experience so as to do better in the future. These are usually time sequential steps in task performance" \cite{doi:10.1002/9781118131350.ch34}. Leveraging our understanding and experience in Machine Learning, we question who is the entity that takes the supervisor role.

\begin{figure}[htbp]
    \centering
    \begin{subfigure}[b]{0.48\linewidth}
        \includegraphics[width=\textwidth]{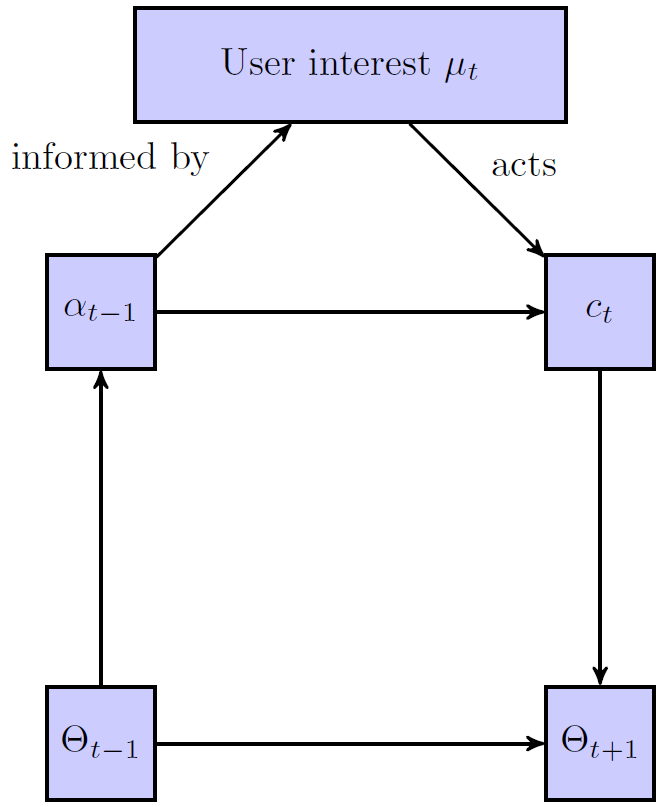}
        \caption{User Supervisor}
        \label{fig:reformulation_figure_a}
    \end{subfigure}
    \begin{subfigure}[b]{0.48\linewidth}
        \includegraphics[width=\textwidth]{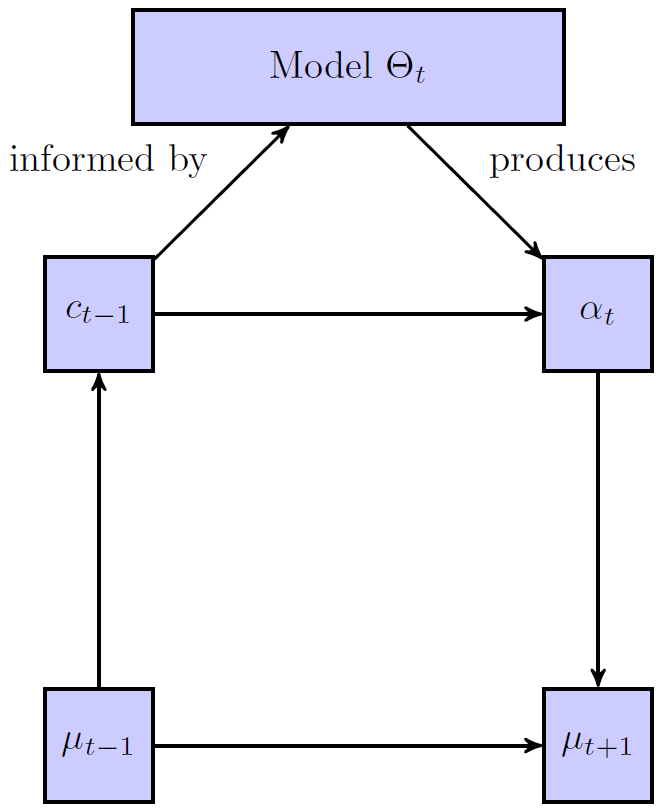}
        \caption{AI Model Supervisor}
        \label{fig:reformulation_figure_b}
    \end{subfigure}
    \caption{A theoretical analysis of the interaction dynamics between a single user and a Recommender System model.}
    \label{fig:reformulation_figure}
\end{figure}

Figure \ref{fig:reformulation_figure_a} is a reformulation in time, of the interaction model in Figure \ref{fig:setup_figure_b}. At time step $t$, the user has interest $\mu_t$ which is inevitably influenced by the recommended list of items at the previous time step $\alpha_{t-1}$. This follows from the edge ($\alpha_t$, $\mu_{t+1}$) in Figure \ref{fig:setup_figure_b}. The user acts on making a choice following or not following the recommendation - represented by the edge ($\mu_t$, $c_t$). All other edges in the user supervisor model follow from the interaction model in Figure \ref{fig:setup_figure_b}. The model $\Theta_{t-1}$ influences the set of items that are shown to the user $\alpha_{t-1}$. The user's actions, the previous version of the model, as well as the previously recommended items will influence $\Theta_{t+1}$. 
The $\Theta_t$, $\Theta_{t+1}$ layer represents the "Computer" as introduced in the human supervisory control model (Figure \ref{fig:setup_figure_a}), and that layer is connected to the user through the "Display" and "Controller" nodes - the list of recommendations and user's feedback respectively. 

Figure \ref{fig:reformulation_figure_b} shows another reformulation of the same interaction model, however the model is seen as the supervisor. At time step $t$, $\Theta_{t}$ is influenced by the set of revealed user preferences $c_{t-1}$, this follows from the edge ($\alpha_t$, $\Theta_{t+1}$) in Figure \ref{fig:setup_figure_b}. A new set of recommended items is produced and presented to the user $\alpha_t$. User's interest $\mu_{t-1}$ influence their actions $c_{t-1}$ as well as their interest in the next time step $\mu_t$. The user's actions, their previous interest, and the previously recommended items will all influence $\mu_{t+1}$. 
The user interest $\mu_t$, $\mu_{t+1}$ layer represents the "Computer" as denoted the human supervisory control model (Figure \ref{fig:setup_figure_a}), and that layer is connected to the AI model through the "Display" and "Controller" nodes - the user's feedback and the list of recommendations respectively.

Based on this analysis we see that both of these scenarios are an adequate representation of the original interaction model introduced by \citeauthor{Jiang:2019:DFL:3306618.3314288}. Both the user, as well as the AI model, could take the role of the supervisor in the sense introduced by the worldview of human-in-the-loop and Supervisory Control system design.

\section{Evolving Evaluation Metrics}
In this section we briefly introduce RSs performance evaluation metrics in order to identify a gap which we think is important to address. We follow by developing a metric we call a \textit{barrier-to-exit} and demonstrate how it could be applied practically. Our goal is to describe the need for it conceptually and invite researchers and practitioners to help us address that need.

Broadly speaking there are two main types of evaluation metrics - metrics that evaluate algorithmic outputs and those related to quantifying human behavior (for example using click rate as a proxy for user preference). We elaborate the need for evaluation metrics that analyze the interactions between people and algorithmic systems. \citeauthor{herlocker2004evaluating} have developed a taxonomy of Collaborative Filtering RS evaluation metrics collapsed into three equivalence classes - predictive accuracy metrics, classification accuracy metrics, and rank accuracy metrics. To the best of our knowledge, those metrics fail to consider mutability and the dynamics of changing user preference. According to multiple studies \cite{amatriain2009like,amatriain2009rate,hill1995recommending}, user feedback and preferences are inconsistent - users may rate the same item differently at different points in time. Less well-studied are the reasons for these inconsistencies and behavioral aspects of using RSs, including anchoring effects \cite{Zhang:2011:AER:2043932.2044010}, the implications of bias \cite{DBLP:journals/corr/abs-1811-01461} which may lead to contamination of the inputs of the RSs, artificially improving accuracy scores, and providing ways for the model to manipulate the system in order to optimize certain quantities. \citeauthor{10.1007/978-3-642-31454-4_20} have derived a metric they call a magic barrier - the lower bound on the root-mean-square error that can be attained by an optimal RS \cite{10.1007/978-3-642-31454-4_20,Krishnan:2014:MLA:2645710.2645740}. The magic barrier marks the point at which user accuracy cannot be enhanced due to the noise and inconsistencies in the data, any further enhancement will lead to overfitting. 

\citeauthor{schmit2017human} propose to measure RS performance by adapting the notion of regret from the multi-armed bandit literature. In the traditional contextual bandit literature, in order to analyze the behavior of a user (referred to as an agent or a forecaster), we compare its performance with that of an optimal strategy, that consistently plays the arm that results in the most reward. Then the notion of regret roughly refers to the difference between the optimal strategy and the agent's strategy in terms of reward. The formalization of this concepts allows us to distinguish between two notions of averaged regret - expected regret and pseudo-regret, and bound the pseudo-regret within a logarithmic function of the time horizon (number of time steps we're interested in) \cite{DBLP:journals/corr/abs-1204-5721}. \citeauthor{schmit2017human} have done extensive simulations to study how to incentivize exploration from the perspective of the algorithm, referred to as a social planner who is interacting with self-interested decision-makers. In our research and work we argue that this view is helpful but very difficult to apply in practical settings due to our inability to create reward functions which are capable to approximate users' real-world preference. We therefore see a need for the development of RSs evaluation metrics which take a human-centered point of view vs and algorithmic-centered point of view.

RSs researchers and developers have worked on incorporating the notions of diversity, popularity and serendipity in the evaluation of RS models \cite{Nguyen:2018:UPU:3301908.3301921}. \citeauthor{Nguyen:2018:UPU:3301908.3301921} have done that through qualitative studies where users were asked to measure on a 5-point scale:
\begin{itemize}
\item (Diversity) How dissimilar are the movies in the list from each other?
\item (Popularity) How popular are the movies in the list?
\item How surprised are you to see these movies being recommended to you?
\end{itemize} 
We build on that work and perform multiple experiments that aim to help RSs researchers and developers frame the notion of mutability and human self-determination in the way users interact with a RS model.

We argue that the AI System evaluation metrics should not treat data as static artifacts, instead, as in the case of identity transitions, data should be seen as a part of a dynamic and multifaceted process. For example, in gender transitions, "there's rarely a linear path from one gender to another - it instead may include pauses, regression and tangents - and identity continuous to  adjust post-transition" \cite{Haimson:2016:DFC:2858036.2858136}. Moreover, we need hybrid evaluation metrics that operate on the level of human-algorithmic interaction. The quality and availability of that data will inevitably depend on the design of the AI system. As explored by \citeauthor{schnabel2018improving}, "changes in the user interface can impact the quality and quantity of user feedback - and therefore the learning accuracy" \cite{schnabel2018improving}. We argue that implicit and explicit feedback data, which is part of what we call human-algorithmic interaction data, need to be carefully considered as it cannot be separated from the design choices which were employed in the development of the systems that produced them. Furthermore, rather than focusing only on output oriented metrics such as accuracy, we need to develop metrics which understand what is the user's true preference as a function of the human-algorithmic system.   

\section{Barrier to Exit}
We start by recognizing that a user's identity, self identification, and their preferences are ultimately impossible to quantify. As studied by \citeauthor{Haimson:2016:DFC:2858036.2858136}, identities are socially constructed \cite{berger1991social,Mayer-Schonberger:2009:DVF:1804518} and identity transitions are social \cite{Haimson:2015:OIE:2702123.2702270,Haimson:2015:DSS:2675133.2675152,haimson2016constructing}. The field of Social Constructionism within Critical Sociology provides us with important perspectives on mutability. If we show that a certain phenomenon that we are trying to measure (directly or through proxies) is socially constructed than it follows that it could be constructed differently, which shows that it is mutable. In its most extreme form, everything becomes a social construct, and there is nothing else we can know of the world. However, such a view undermines the reliability of all knowledge claims and all ethical claims itself and therefore it also undermines its own knowledge claims and ethical claims. Therefore not all concepts are subject to Social Constructionism \cite{ElderVass2012TowardsAR}.

In what follows we aim to analyze identity transitions on a more granular level - What is the effort that a user needs to expend in order for the system to recognize their change in preference? We set out to create a conceptual model where we are able to quantify that effort for specific categories and call it a \textit{barrier-to-exit} (BtE) from each respective category of items. The \textit{barrier-to-exit} for a RS as a whole we define as the average of the BtE scores for all represented categories. We acknowledge the complexity of measuring this concept as, if we are to fully analyze it, we'd need to have the possibility to play out scenarios in counterfactual worlds. To develop one possible proxy to the BtE, we develop a metric that incorporates the change in user's revealed preferences over time.

\begin{figure}[htbp]
    \centering
    \begin{subfigure}[b]{1\linewidth}
        \includegraphics[width=\textwidth]{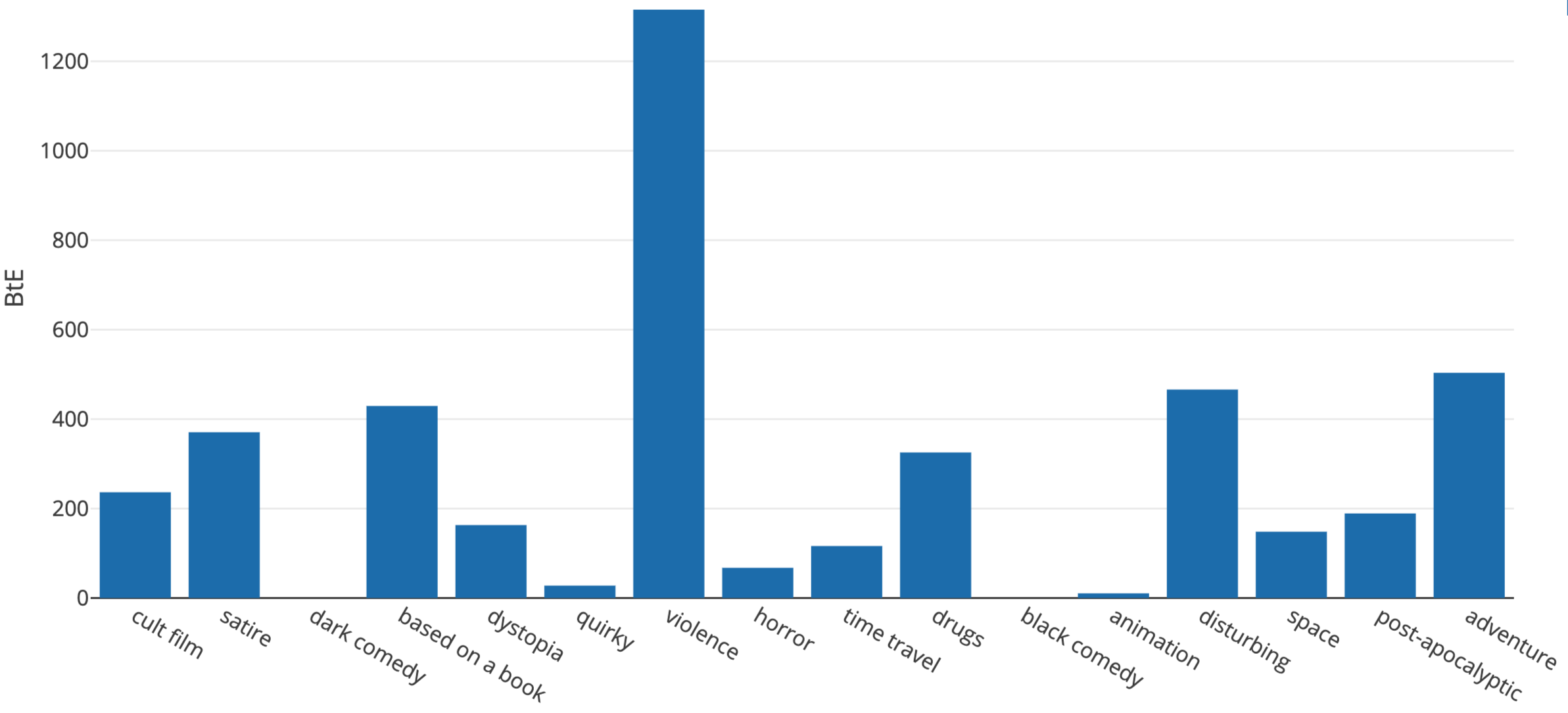}
        \caption{Per category $BtE$ scores}
        \label{fig:results_figure_a}
    \end{subfigure}
    \begin{subfigure}[b]{1\linewidth}
        \includegraphics[width=\textwidth]{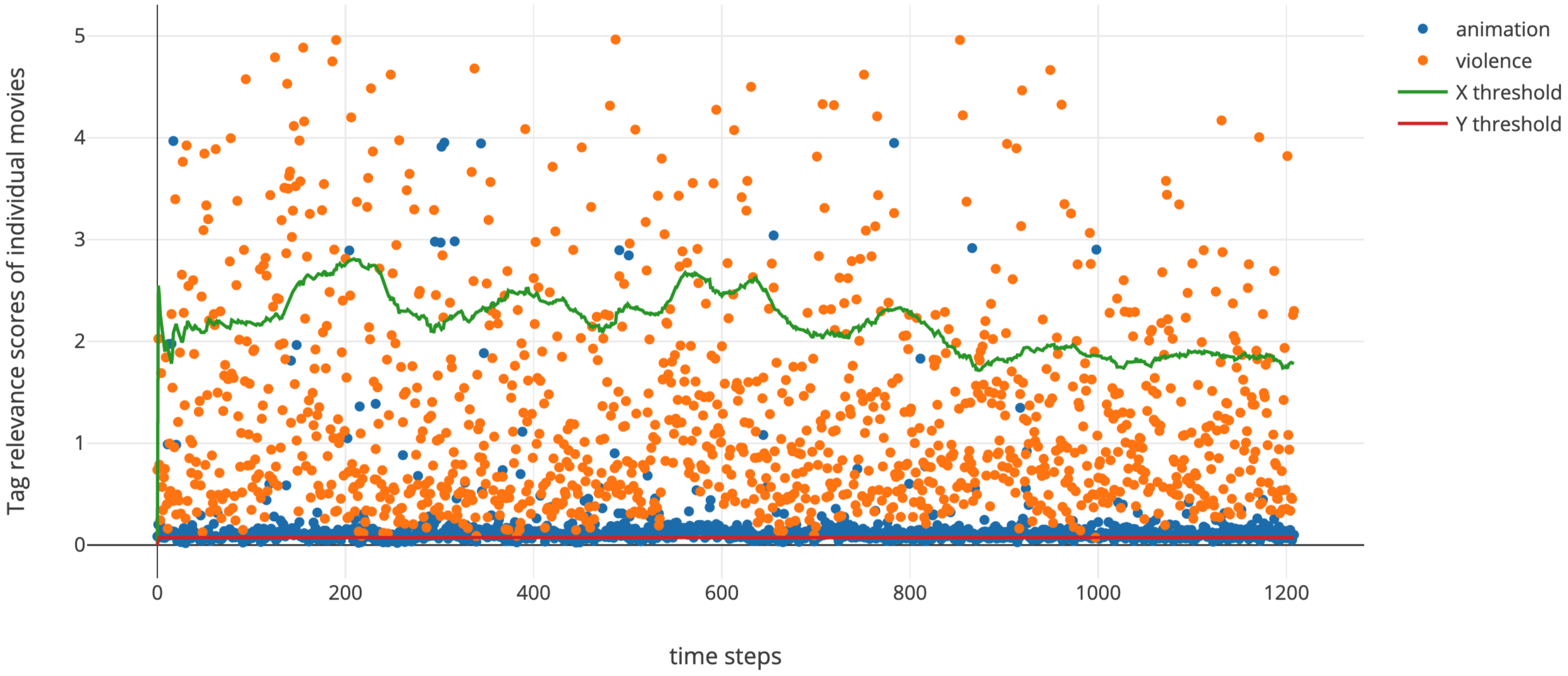}
        \caption{Raw movie ratings data for specific categories}
        \label{fig:results_figure_b}
    \end{subfigure}
    \caption{The \textit{barrier-to-exit} from specific categories of movies - considering the user's raitings of specific movies within a category, weighted by the movies' relevance score to that category.}
    \label{fig:results_figure}
\end{figure}

We carry out our research experiments on MovieLens's 1M dataset \cite{grouplens}. MovieLens is a well-known movie RS that has been in continuous use since 1997 with more than 200,000 users providing more than 20 million ratings. Since 2003, it uses a well-known and widely used recommendation algorithm based on item-item collaborative filtering. The platform provides a \textit{"Top Picks for You"} feature which shows users recommended movies they have not seen ordered accordingly to their prediction ratings. The recommendation lists data is not publicly available and therefore we were not able to incorporate it into our experiments. The publicly available data consists of metadata about the list of movies users have rated - timestamp, headline, and rating (on a scale from 0.5 to 5), as well as a user-generated dataset called \textit{tag-cloud} which gives us a list of tags (short phrases) associated with specific movies. We use the \textit{tag-cloud} data to extract a set of finite or countably infinite possible categories $\kappa_1$,... ,$\kappa_l$ $\in$ $K$, where $l$ is the number of categories. At every time step $t$ the RS serves the user the \textit{Top Picks for You} set of recommended items. We denote $c_t$ to be the user feedback in the form of the set of movie ratings the user has given at that time step: $c_t$=($c_t^1$,...,$c_t^l$) $\in$ ${\rm I\!R}^l$, where $c_t^i$ represents the user's revealed preference in category $i$ at time step $t$. If $n$ is the number of movies the user has rated at time step $t$, we denote the rating received for those movies as ($r_{t1}$,...,$r_{tn}$), and the tag-relevance score between each of the movies and the tag(category) $\kappa_i$ as ($m_{t1}^i$,...,$m_{tn}^i$). We calculate the notion of user feedback or revealed preferences in the following way:
\begin{align}
c_t^i = \sum_{j=1}^{n} m_{tj}^i * r_{tj}
\end{align}
We extract an upper($X$) and lower($Y$) interaction thresholds per category at each time step, considering a past time horizon of length $\nu$:
\begin{align}
X_t^i = mean(c_{t-\nu}^i,...,c_t^i) + 2 * std(c_{t-\nu}^i,...,c_t^i) \\
Y_t^i = mean(c_{t-\nu}^i,...,c_t^i) - 2 * std(c_{t-\nu}^i,...,c_t^i) 
\end{align}
Let $X_t$ and $Y_t$ be the averaged values of the thresholds across all categories $i$ $\in$ $K$. We are interested in the time steps where the user feedback is above the upper threshold $X_t$ and those where it is below the lower threshold $Y_t$, for $t$=1,... We denote a time step where these cases have occurred as $t_x$ and $t_y$ respectively. This would mean that:
\begin{align}
Y_\tau < c_\tau^i < X_\tau \text{, for } \tau \in (t_x, t_y) 
\end{align}
For the time period ($t_x$, $t_y$) we denote the \textit{barrier-to-exit} $BtE$($t_x$, $t_y$, $i$) to be the area under the curve defined by ($c_{t_x}^i$,...,$c_{t_y}^i$) which in the discrete case is the sum $\sum_{\tau=t_x}^{t_y} c_\tau^i$. The mean of all available $BtE$($t_x$, $t_y$, $i$) within a certain time frame we define as the overall \textit{barrier-to-exit} $BtE^i$ for category $i$. Figure \ref{fig:results_figure_a} plots the \textit{barrier-to-exit} derived through this methodology for the categorical tags: \textit{cult film}, \textit{satire}, \textit{dark comedy}, \textit{based on a book}, \textit{dystopia}, \textit{quirky}, \textit{violence}, \textit{horror}, \textit{time travel}, \textit{drugs}, \textit{black comedy}, \textit{animation}, \textit{disturbing}, \textit{space}, \textit{post-apocalyptic}, and \textit{adventure}, for an individual user with user id = 3841, belonging to the MovieLens's 1M dataset \cite{grouplens}. Figure \ref{fig:results_figure_b} shows the interaction thresholds $X$ and $Y$, as well as the raw movie ratings the user has given for movies labeled with tags \textit{violence} and \textit{animation}. 

We have no way to know the user's true preference, but only have access to longitudinal data of movie ratings. Still, we propose that a RS which has a lower overall $BtE$ score would more readily allow for the user to change their preference at their own will. Moreover a human-centered RS design could continuously monitor the per user per category $BtE$ scores and provide the user additional explanatory information when certain thresholds are reached. Leveraging the body of work in Explainability and Interpretability of AI Systems, the \textit{barrier-to-exit} metric could be used as a way for the users themselves to monitor the behavior of the AI System (and not the other way around). The RS could allow for the user to monitor and adjust the interaction thresholds according to their true preference. By incorporating these design recommendations, the RS creators could engineer negative feedback loops that provide the needed checks and balances and signal the user when those thresholds have been reached.  

\section{Discussion and Future Work}
When interacting with technology, we cannot change or reverse the digital artifacts that we leave - we cannot unclick a link or unwatch a movie. For individual users, the digital artifacts link the past and the present in ways that may be problematic when we want to actively separate ourselves from our past. As explored by \citeauthor{Haimson:2016:DFC:2858036.2858136}, system design which supports the separation between users' past and present identities, through a form of "intentional forgetting" \cite{Haimson:2016:DFC:2858036.2858136}, could allow for mutability and benefit those who have decided to make identity transitions. We go on to propose that the \textit{barrier-to-exit} metric described here could give users more visibility and control over the impact that an AI System has on their true user preferences. 

As shown by \citeauthor{milano2019recommender}, we need an overarching framework to help us "reconstruct the whole debate, understand its main issues, and hence offer a starting point for better ways of designing RS and regulating their use" \cite{milano2019recommender}. 
A collaborative and multi-stakeholder approach could allow us to distribute agency across multiple human and non-human actors \cite{bruni2007reassembling}. In the case of RSs, this would include the people creating and publishing the content being recommended, the people who are users of the platform, the people who are part of dynamic and likely ongoing data annotation process, the developers and maintainers of the RS itself, the RSs model, the people and organizations who are consuming the digital traces produced by the users of the system for other purposes, and others. 

Going beyond RSs, our research could meaningfully contribute to the field of investigating the implications and trustworthiness of data and AI models. An evolution of the \textit{barrier-to-exit} concept and metric may give us an axis on which to compare and study the interactions between pre-existing social categories, in that sense it aims to be intersectional. As \citeauthor{doi:10.1080/1369118X.2019.1573912} points out intersectionality \cite{carastathis2016intersectionality,mitchell2018prediction} "is concerned with both the production and hierarchical ordering of identity categories" \cite{doi:10.1080/1369118X.2019.1573912} such as race, gender, and many others. Looking at the interactions between pre-existing categories gives us new insights about data and algorithmic disparities, as demonstrated by \citeauthor{pmlr-v81-buolamwini18a} in their work on facial recognition systems.

There are ripple effects to our interactions with recommendation engines - our data goes on to be used in other settings as raw data for other algorithms. Algorithmic assemblages, operating in "networked information environments" \cite{ananny2018seeing} do not only produce individual distributive outcomes but are also "intimately bound up in the production of particular kinds of meaning, reinforcing certain discursive frames over others" \cite{doi:10.1080/1369118X.2019.1573912}. Addressing these considerations requires employing frameworks that (1) allow us to think in terms of algorithmic assemblages situated within a sociotechnical context \cite{selbst2019fairness,cinnamon2017social,doi:10.1080/1369118X.2016.1200645} , and (2) see data and algorithms not as static artifacts but as part of an ongoing process which evolves at different time-scales \cite{overdorf2018questioning,liu2018delayed,ananny2018seeing}. Specifically, (1) we need practitioners and researchers to employ evaluation metrics that go beyond optimizing for the inputs and outputs of a single algorithm, and (2) consider how the data that an algorithm operates on is part of a dynamic larger whole (for example user's preference or identity). Building on prior work showing the value of interdisciplinary research, we see an urgent need to draw from the field of Philosophy, Systems Theory, Complexity Theory, and others, which have developed the concepts of emergence and synergy. "Synergy is the only word in our language that means behavior of whole systems unpredicted by the separately observed behaviors of any of the system's separate parts or any subassembly of the system's parts. There is nothing in the chemistry of a toenail that predicts the existence of a human being" \cite{fuller2008operating}. Similarly, its imperative for us to question the predictive power of individual datums, for examples click-rate, over a person's political inclination or other aspects of their identity.

\section{Conclusion}
Defining the concepts of human self-determination, mutability, and identity, in the language of AI Systems is challenging. It requires us to take an interdisciplinary and longitudinal approach to research. In this work, we provide a theoretical analysis of the human-algorithmic interaction between a single user and a RS model in time, building on prior work by \citeauthor{Jiang:2019:DFL:3306618.3314288} and the fields of Supervisory Control within Human Factors and Ergonomics, and Science and Technology Studies. We show that there's a need for evaluation metrics which operate on human-algorithmic interaction data and consider mutability and identity transitions. We propose a new metric (\textit{barrier-to-exit}) and demonstrate how it can be used in the context of the MovieLens Recommender System. Going beyond movie recommendations, we identify key areas where improvements in the design of RSs could allow for their users to more readily change and have agency over their true user preference.
Our work has two main limitations. First, the \textit{barrier-to-exit} metric proposed here cannot be easily generalized across multiple users but instead is meant to be used on the level of an individual user. At the same time, it is a metric that measures the ability of a RS model to allow for mutability in user preference and therefore it is meant to bring visibility and inform multiple stakeholders including the AI System designers. Second, the list of social categories that is used when deriving the metric is not fixed but will differ in different cases. Future work needs to further our understanding of the interaction between social categories.





\bibliography{library}
\bibliographystyle{aaai}

\end{document}